\documentclass[final,5p,times,twocolumn,authoryear]{elsarticle}

\usepackage{amssymb}
\usepackage{lipsum}
\usepackage{multirow}
\usepackage{color}
\usepackage{hyperref}

\journal{High Energy Astrophysics}

\begin{document}
\begin{frontmatter}

\title{Seyfert galaxy targets for KM3NeT neutrino telescope}

\author[apc,epfl]{A. Neronov}
\author[apc,bitp,kau]{D. Savchenko}
\author[bitp]{M. Poleshchuk}

\affiliation[apc]{organization={Université Paris Cité, CNRS, Astroparticule et Cosmologie},
            postcode={F-75013}, 
            city={Paris},
            country={France}}
\affiliation[epfl]{organization={Laboratory of Astrophysics, École Polytechnique Fédérale de Lausanne},
            postcode={CH-1015}, 
            city={Lausanne},
            country={Switzerland}}
\affiliation[bitp]{organization={Bogolyubov Institute for Theoretical Physics of the NAS of Ukraine},
            postcode={03143}, 
            city={Kyiv},
            country={Ukraine}}
\affiliation[kau]{organization={Kyiv Academic University},
            postcode={03142}, 
            city={Kyiv},
            country={Ukraine}}

\begin{abstract}
Neutrino signal from a population of Seyfert galaxies has been detected by IceCube neutrino telescope in the muon neutrino channel that has sensitivity mostly to the Northern Hemisphere sources. This detection can be verified by KM3NeT telescope that has sensitivity also in the Southern Hemisphere. We define a catalog of Seyfert galaxies that are expected to be detectable with KM3NeT, assuming that the neutrino luminosity scales with the intrinsic hard X-ray luminosity of the sources. We find that four sources: NGC~1068, NGC~4151, NGC~4945 and Circinus galaxy, are detectable by KM3NeT, if their spectra follow either NGC~1068 or NGC~4151 spectral template based on IceCube data.  We discuss uncertainties of the neutrino flux estimate, considering  the Compton-thick nature of three of the four detectable sources: NGC~1068, NGC~4945 and Circinus. The limited catalog of the four sources can be used in KM3NeT source search to reduce the trial factor of analysis aimed at independent verification of the neutrino signal from Seyfert galaxies.    
\end{abstract}


\end{frontmatter}

\section{Introduction}
\label{introduction}

Seyfert galaxies have been recently found to be very-high-energy (VHE) neutrino sources by the IceCube neutrino telescope~\citep{2022Sci...378..538I,neronov2024,2025ApJ...981..131A,2025ApJ...988..141A}. They may provide a sizable contribution to the overall astrophysical neutrino flux in this energy band~\citep{seyfert-theory}. This is surprising, because previous to the IceCube detection, Seyfert galaxies have been considered to be ``high-energy-quiet'' subclass of Active Galactic Nuclei (AGN) that have weak or absent jets and do not show signatures of high-energy particle acceleration. 

High-energy particle interactions leading to production of neutrinos also generate  $\gamma$-rays in the same energy range as neutrinos.  The lack of   $\gamma$-ray flux comparable to the neutrino flux from Seyfert galaxies  suggests that neutrinos are produced in compact dense environment in which $\gamma$-rays are  reprocessed by electromagnetic cascades. Such dense regions exist close supermassive black holes powering the AGN activity~\citep{2020PhRvL.125a1101M,2021ApJ...922...45K,2022arXiv220702097I,2022ApJ...939...43E,2025PhRvD.112l3016B,seyfert-theory}. \cite{seyfert-theory} have suggested that the electromagnetic power injected by high-energy proton interactions ultimately contributes to hard X-ray  luminosity of the source. In this case, the neutrino luminosity of Seyfert galaxies is expected to scale with the  X-ray luminosity of the accretion flow and hence the neutrino luminosity can be estimated from the hard X-ray luminosity. However,  the X-ray emission from compact regions around the supermassive black hole can be blocked by obscuring matter, so that the X-ray luminosity originating from the direct vicinity of the black hole is not always directly measurable.  

Testing the hypothesis of relation between VHE neutrino and X-ray power requires studying the properties of neutrino and X-ray signals from a large number of Seyfert galaxies. This is currently not possible because of the limited sensitivity and partial sky coverage of the IceCube neutrino telescope. Due to the geographical position of the IceCube detector on the South Pole, source detections in the ``track'' channel (which is more sensitive to point sources, compared to an alternative ``cascade'' channel) are  effectively restricted to the Northern sky. 

The new KM3Net neutrino telescope, currently under construction in the Mediterranean Sea, provides major improvement of sensitivity in the Southern sky owing to its location in the Northern Hemisphere~\citep{2024EPJC...84..885K}. This provides an opportunity for independent verification of IceCube detection of neutrinos from Seyfert galaxies and for testing of the hypothesis of the link between neutrino and hard X-ray power. Overall, the sensitivity of KM3NeT is comparable to that of IceCube and detection of individual sources are expected to be moderately significant. In such conditions, KM3NeT data analysis will have to rely on the source searches using pre-defined source catalogs, to reduce the trial factor related to the ``look elsewhere'' effect.  

In what follows we explore the prospects for detection of Seyfert galaxies with KM3NeT and work out a catalog that can be used as the ``pre-defined source list'' in the search for neutrino signal from Seyfert galaxy source class.

\section{Source selection}
\label{sourceselection}

\begin{table*}
\centering
\begin{tabular}{lcccccccl}
Name& RA & Dec & D   & $F_{hX}$ & $L_{hX0}$ & $F_{\nu_\mu,0.3-100}$& $N_H$ &Type \\
 & deg & deg & Mpc & $10^{-11}$ erg/(cm$^2s)$ & $10^{43}$~erg/s & $10^{-11}$~TeV/(cm$^2$s) & $10^{24}$~cm$^{-2}$ & \\
\hline

NGC~1068 & 40.670 & -0.013 & 16.3 &  3.79 & 0.94-22$^a$  & 0.16-30.5 &$\gtrsim 10^a$& Sy2\\
NGC~4151 &  182.636 & 39.406 & 14.2 & 61.89 & & 2.6 & 0.08$^c$&Sy1\\
NGC~4945 & 196.365 & -49.47 & $3.8^h$ & 28.21 & 0.12-1.3$^d$ &1.18-33.2 &2.5-5.3$^d$   & Sy2 \\
Circinus Galaxy & 213.292 & -65.342 & $4.2^h$ & 27.31 & 0.14-0.5$^f$ &1.2-10.4 &6-10$^f$ &  Sy2 \\
\hline
NGC~1106 & 42.671 & 41.67 & 59.5 & 1.9 & 4.09-42.85$^g$ & 0.06-4.5 &4.8$^g$ &Sy2\\
NGC~1142 & 43.8 & -0.184 &118.6 & 8.8 & 35.6-45.9$^g$ & 0.35-1.2 &1.6$^g$ & Sy2\\
NGC~3281 & 157.967 & -34.854 & 43.603 & 8.6 & 5.3-6.5$^g$ & 0.34-1.3 &2.0$^g$ & Sy 2\\
NGC~2110 & 88.046 & -7.457 & 34.5 & 32.89 & & 1.4 &0.04$^b$ & Sy2 \\
NGC~3079 &  150.491 & 55.680 & 15.9 & 3.67 & 0.3-1.7$^b$  &0.15-2.54 & 3$^b$ &Sy2\\
NGC~4388 &  186.445 & 12.662 & 36.2 & 27.89 &  & 1.2 &0.47$^b$ &Sy2 \\
IC4329A  & 207.329 & -30.306 & 69 & 26.32 & & 1.1 & 0.006$^e$ &Sy1 \\
\hline
\end{tabular}
\caption{Seyfert galaxies with expected muon neutrino fluxes in excess of $10^{-11}$~TeV/(cm$^2$s) in the energy range above 300~TeV. $F_{hX}$ and ``Type'' are 15-195~keV flux and source type from Swift/BAT catalog \citep{SWIFTBAT}. $N_H$ is the hydrogen column density measurements collected from literature:   $^a$\citet{2016MNRAS.456L..94M}; $^b$\citet{2017PhDT.......388B}; $^c$\citet{FischerCatalog}; $^d$\citet{2014ApJ...793...26P}; 
$^e$\citet{Ogawa:2019jqg}; $^f$\citet{circinus}; $^g$\citet{Georgantopoulos:2024ggg}. $L_{hX0}$ is the intrinsic hard X-ray luminosity estimates for sources with $N_H>10^{24}$~cm$^2$ from the same references as the $N_H$ estimates. The distances of the two local group sources are from 
$^h$\citet{2009AJ....138..323T}.  Sources listed in the top section of the table are detectable by KM3NeT. Sources in the bottom section are not detectable unless their spectra are harder than those of Seyfert galaxies reported by IceCube. }
\label{tab:galaxies}
\end{table*}

To work out the list of Seyfert galaxies that have the best chance to be detected by KM3NeT,  we extend the approach that has been used by~\cite{neronov2024} for detection of neutrino signal from Seyfert galaxy population with IceCube. We define a catalog of the Seyfert galaxies that are detectable within a very long (twenty years) operation time span of KM3NeT. We estimate the neutrino fluxes adopting the hypothesis that they scale with the intrinsic hard X-ray source luminosity. For the sources with the estimated neutrino flux exceeding KM3Net sensitivity, we predict the exposure time required to reach a significant detection assuming spectral models considered by~\cite{seyfert-theory}.

\begin{figure}[t]
\includegraphics[width=\linewidth]{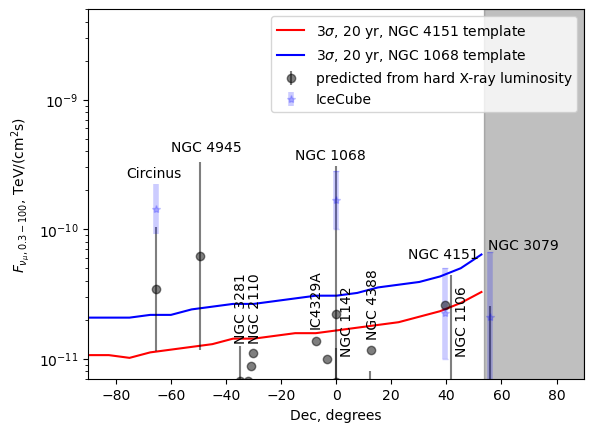}
\caption{Expected neutrino flux from Seyfert galaxies in $0.3-100$~TeV energy range, estimated from Eq.~\ref{eq:fnu}.  Vertical bars indicate uncertainties derived from uncertainties of measurements of intrinsic luminosity for Compton-thick sources. Red and blue lines show  KM3Net $3\sigma$ detection thresholds for 20 years of exposure for two different spectral templates discussed in the text. Gray shading shows the visibility constraint of KM3NeT in the track observation mode.}
\label{fig:south_sources}
\end{figure}

The sources  are selected from the Swift-BAT 105-month source catalog~\citep{SWIFTBAT} which reports the hard X-ray flux of sources in $15-195$~keV energy range. We complement the information in the catalog with the measurements of the hydrogen column density $N_H$ and, where necessary, with information on the distance to the source. This allows us to determine relation between the observed flux and intrinsic source luminosity.  To estimate the intrinsic hard X-ray luminosity, we assume that for sources with $N_H<10^{24}$~cm$^{-2}$, the photoelectric absorption in $15-195$~keV band is neglibible or moderate and the X-ray flux in this band,  $F_{hX}$, directly provides a good estimate of the source luminosity $L_{hX}=4\pi D^2 F_{hX}$, where $D$ is the distance to the source. A limitation of such estimate is that it implies that the reflected component of the X-ray flux is sub-dominant compared to the direct component.  For sources with $N_H>10^{24}$  we have searched literature for more detailed estimates of the intrinsic source luminosity $L_{hX0}$, derived through modeling of the hard X-ray emission spectra with account of absorption and reflection. In case the intrinsic luminosity is given in an energy range different from $15-195$ keV, we re-calculate the luminosity into this energy range.

\begin{figure*}
\includegraphics[width=0.33\linewidth]{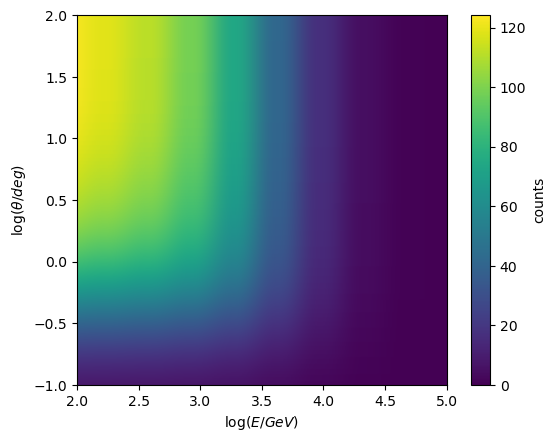}
\includegraphics[width=0.33\linewidth]{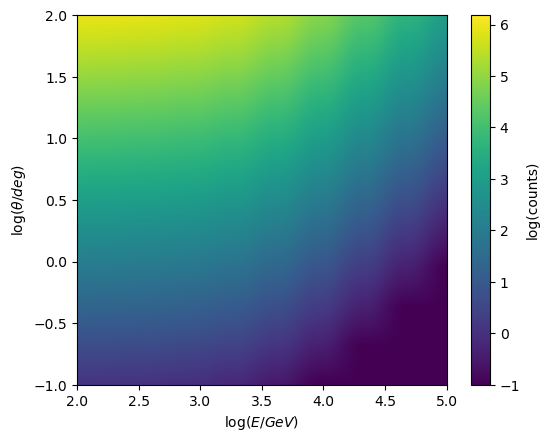}
\includegraphics[width=0.33\linewidth]{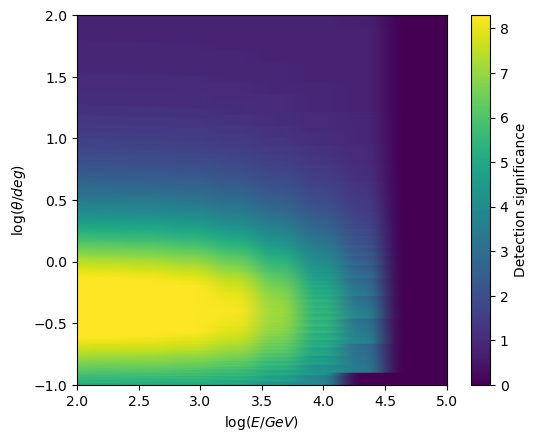}
\caption{Left: cumulative source counts for NGC~1068 ($\epsilon=1$ template) as a function of energy threshold $E$ and radius of the signal extraction circle $\theta$. Middle: cumulative background counts in the same representation. Right: Detection significance in the same representation.}
\label{fig:counts}
\end{figure*}

Estimates of the intrinsic hard-X-ray luminosity for sources with large hydrogen column density suffer from a range of uncertainties. One source of uncertainty stems from large number of parameters of models used to describe the sources. These parameters include the geometry of the accretion flow that is often not well constrained by the observational data. Considering this fact, we 
always keep the lower bound on the intrinsic source luminosity $L_{hX0}\ge L_{hX}$ and estimate the range of $L_{hX0}$ from the span of model estimates reported in the literature.  Another source of uncertainty is the extrapolation of the powerlaw intrinsic source spectrum into 15-195 keV band, in cases when the intrinsic luminosity is given in a different band. Finally, there is still another source of uncertainty related to the source variability. Models of hard X-ray emission relying on e.g. \mbox{NuSTAR} data report analysis of snapshot observations at a given moment in time, while the Swift/BAT measurements of the flux refer to time-average source flux. Hard X-ray emission from Seyfert galaxies is known to be variable, and it is not clear if the snapshot data are representative for the time-average source state.  

Following \cite{neronov2024}, we adopt a hypothesis that the neutrino luminosity of the source in TeV energy range is determined by its intrinsic hard X-ray luminosity, with the proportionality coefficient inferred from a sample of 3 sources (NGC~1068, NGC~4151 and NGC~3079) for which neutrino flux estimates are available. We adopt the following scaling for the all-flavour neutrino flux in IceCube energy range 0.3-100~TeV:
\begin{equation}
\label{eq:fnu}
    F_{\nu, 0.3-100} \sim 0.2 \cdot \frac{L_{hX0}}{4 \pi D^2}
\end{equation}
The muon neutrino flux is 1/3 of the all-flavor  flux $F_{\nu_\mu,0.3-100}=F_{\nu, 0.3-100}/3$, assuming full mixing of neutrino flavors during propagation from the source to the Earth. 

We concentrate our attention on sources with estimated  muon neutrino flux in excess of $10^{-11}$~TeV/(cm$^2$s), which is at the limit of detectability at $3\sigma$ significance level for both IceCube and KM3NeT. Table \ref{tab:galaxies} gives the list of sources passing the selection conditions. Fig. \ref{fig:south_sources} provides a visual representation of the expected neutrino flux levels from the sources in Table \ref{tab:galaxies}.

\section{Comparison with KM3NeT sensitivity and visibility constraints. }

Not all sources listed in Table \ref{tab:galaxies} are detectable with KM3NeT. KM3NeT is sensitive in the ``track'' detection channel only to sources at zenith angle larger than 90 degrees. One of the sources listed in Table \ref{tab:galaxies}, NGC~3079, is situated in the part of the sky that is always below the 90 degree zenith angle for KM3NeT that is situated as the latitude $36.27^\circ$ North. Thus, it is not observable in the ``track'' detection channel. 

Sources within the visibility limit of KM3NeT need to be above the sensitivity limit to be detectable. To estimate the sensitivity of the instrument,  we use the open data release of the KM3NeT/ARCA230 instrument response functions reported by~\citet{km3net_2024_10955399}. In these response functions, the effective collection area is a function of zenith angle and background is reported as a function of declination.  

We use the response functions to model the neutrino signals from sources with spectra considered by  \cite{seyfert-theory}. These spectra are relatively well constrained in the energy range of sensitivity of IceCube, but they suffer from large (order-of-magnitude)  uncertainties at energies below TeV. \cite{seyfert-theory} provide templates for NGC~1068 and NGC~4151 spectral models with the overall neutrino luminosities reaching $\epsilon=0.1$ and $\epsilon=1$ of the intrinsic hard X-ray luminosity of the source. 
For each spectral template, we convolve the neutrino spectrum with the instrument response functions at the source declination, to predict the number of track event counts as a function of the reconstructed energy $E$ and distance from the source $\theta$. Left panel of Fig. \ref{fig:counts} shows an example of the model prediction for NGC~1068 neutrino signal for 20~year exposure for the spectral template with $\epsilon=1$.  Middle panel of the figure shows the background atmospheric neutrino signal at the position of the source, also as a function of the energy threshold $E$ and off-source angle $\theta$. Comparing the source counts $S$ and background counts $B$, we find the detection significance of the source for each choice of $E$, $\theta$ assuming that the overall count statistics $S+B$ follows Poisson distribution. Right panel of Fig. \ref{fig:counts} shows the detection significance for each $E$, $\theta$ choice. 

For each spectral template, we estimate the minimal normalization of the neutrino flux at which the maximal detection significance for any pair of $E,\theta$ is equal to $3$ in a given exposure time. Red and blue lines in Fig. \ref{fig:south_sources} show the $3\sigma$ detection limits for a 20-year operation time span of KM3NeT calculated in this way, as a function of declination. The detection limit worsens with the increasing declination, mostly because of the decreasing effective exposure time as the source mounts over the horizon for more prolonged periods. The NGC~4151 template has harder spectrum compared to the NGC~1068 template. This explains a factor-of-two better detection limit for this template. 

\section{Sources detectable with KM3NeT}

Only four sources listed in the first half of Table \ref{tab:galaxies}: NGC~1068, NGC~4151, NGC~4945 and Circinus galaxy, are found above the detection threshold, even for a very long data taking time span, 20~yr. We list these  sources in the top part of Table \ref{tab:galaxies}. We propose that this restricted source list is used as an input catalog in analysis aimed at confirmation of neutrino signal from Seyfert galaxies with KM3NeT. Sources listed in the bottom part of Table \ref{tab:galaxies} may be detectable if their spectra are harder compared to the templates used in our estimates. Even if these sources would be not detected by KM3NeT, upper limits on their fluxes would still be useful for testing the hypothesis of relation between the hard X-ray and neutrino luminosities and ultimately for calibration of the scaling between neutrino and hard X-ray luminosities (in case the hypothesis is confirmed). 

For the four sources that are presumably detectable within a long KM3NeT exposure, we perform a more detailed analysis to estimate the exposure needed for the source detection. The spectral shapes of the individual sources are not known a-priori and  we explore four different templates proposed by~\cite{seyfert-theory}. For each assumed source spectrum, we generate a number of pseudo-experiments by sampling the events from a Poisson distribution in fixed energy and distance-to-source bins. The expected number of background events in each energy bin $i$ can be obtained from the instrument response functions provided by \citet{km3net_2024_10955399}. The events are distributed proportionally to the area $\theta_j \Delta \theta_j$ of concentric rings of off-source angle $\theta$, so that in each bin $j$ the number of background events $B_{ij}\propto \theta_j\Delta\theta_j$.  The source events are sampled following the spectral templates discussed in the previous sections. For each source we consider the source spectrum to be the template spectrum renormalized by a factor $F_{\nu_\mu, 0.3-100}/F_{\nu_\mu, 0.3-100,ref}$, where $F_{\nu_\mu, 0.3-100,ref}$ is the neutrino flux of the reference source (either NGC~4151 or NGC~1068) and $F_{\nu_\mu, 0.3-100}$ is the flux of the model source of interest. We convolve the source spectrum with the effective area, point spread function and energy response available in the instrument response functions dataset to obtain the expected number of source counts $S_{ij}$ in each energy bin $i$ and off-source angle bin $j$, within an assumed exposure time $T$. The total expected number of events is hence $N_{ij}= S_{ij}+B_{ij}$. In each pseudo-experiment, we generate a realization $s_{ij},b_{ij}$ of the count data, with the total event number $n_{ij}=s_{ij}+b_{ij}$. 

We use the binned likelihood approach to compare the likelihoods of signal-plus-background, 
\begin{equation}
    \log L(\mu) = \sum_\mathrm{bins} n_{ij}\log\left(\mu S_{ij}+B_{ij}\right) - \mu S_{ij} - B_{ij}.
\end{equation}
so that $L(0)$ corresponds to the likelihood of background-only hypothesis and $\mu\ne 0$ are the likelihoods of background+signal hypotheses. 
For each pseudo-experiment, we estimate the signal strength by finding the value of $\mu=\hat\mu$ that maximizes the likelihood. We then compute the Test Statistic value
\begin{equation}
  TS = - 2 \left( \log L(0) - \log L(\hat\mu) \right). 
\end{equation}
For an estimate of the time to $3\sigma$ detection, we require that 50\% of the pseudo-experiments with a given exposure time $T$ result in $TS$ value larger than $9$. For the $5\sigma$ detection, we impose that 50\% of the pseudo-experiments result in $TS>25$.

We explore several benchmark neutrino flux levels for the  two new sources, Circinus galaxy and NGC~4945: the maximal and minimal values reported in  Table \ref{tab:galaxies}. Those provide the minimal and maximal possible time to detection, respectively. For NGC~1068 and NGC~4151 we use IceCube measurement uncertainty instead of the maximal and minimal flux level estimates derived from the hard X-ray data.   

Table \ref{tab:results} shows the results of our calculations. One can see that the time to detection varies in wide ranges, from a fraction of a year up to multi-decade long time scales, depending on assumptions about the source spectral shape and flux normalization.

\begin{table*}[ht]
\begin{tabular}{@{}l|rr|rr|rr|rr@{}}
\hline
 & \multicolumn{2}{c|}{NGC1068 with $\varepsilon_\nu = 0.1$} 
 & \multicolumn{2}{c|}{NGC1068 with $\varepsilon_\nu = 1$} 
 & \multicolumn{2}{c|}{NGC4151 with $\varepsilon_\nu = 0.1$} 
 & \multicolumn{2}{c}{NGC4151 with $\varepsilon_\nu = 1$} \\
 & $3\sigma$ & $5\sigma$ & $3\sigma$ & $5\sigma$ & $3\sigma$ & $5\sigma$ & $3\sigma$ & $5\sigma$ \\
\hline
NGC1068 & $1.6 - 7.4$ & $3.6 - 18.5$ & $0.9 - 4.0$ & $2.0 - 10.1$ & - & - & - & - \\
NGC~4151 & - & - & - & - & $>9.8$ & $>21.0$ & $>7.6$ & $>17.0$ \\
NGC~4945 & $>0.8$ & $>1.8$ & $>0.4$ & $>1.0$ & $>0.4$ & $>0.7$ & $0.2 - 30.0$ & $>0.5$ \\
Circinus Galaxy & $>4.1$ & $>9.3$ & $>2.2$ & $>5.2$ & $>1.6$ & $>2.7$ & $1.1 - 25.0$ & $>2.1$ \\
\hline
\end{tabular}
\caption{Time (in years) to detection with $3 \sigma$ and $5 \sigma$ significance of various sources assuming different spectral templates for neutrino signal. 
Only lower bound is given in cases when lowest predicted flux value wouldn't allow achieving the desirable significance during a realistic instrument lifetime.
}
\label{tab:results}
\end{table*}

\section{Discussion}

Our analysis shows that KM3NeT telescope should be able to scrutinize the IceCube result on detection of neutrino signal from Seyfert galaxies, by (a) confirming the detections of NGC~1068 and NGC~4151 sources reported by IceCube and by detecting two more sources, NGC~4945 and Circinus galaxy, in the Southern Hemisphere. This would be important for establishment of Seyfert galaxies as a confirmed very-high-energy neutrino source class. Given the weakness of the signal from individual Seyfert galaxies, incorporation of the catalog  given in the upper part of Table \ref{tab:galaxies} into  KM3NeT data analysis can improve the sensitivity of the search of the signal. The technique of searching for the signal from a pre-selected source list is commonly used in neutrino astronomy for reduction of the trial factor associated to the ``look elsewhere effect''. Testing the presence of the signal at positions of several  brightest candidate targets reduces this trial factor. 

The source catalog of the upper part of Table \ref{tab:galaxies} includes only two sources in addition to the sources already reported by IceCube. Remarkably, both additional sources accessible to KM3NeT, the Circinus galaxy and NGC~4945, are closer than the Northern Hemisphere sources, they are both members of the Local Group of galaxies that includes the Milky Way. This potentially opens a possibility for localisation of the neutrino source within the galaxy.  Both new sources show the AGN and starburst activities.  These two physical phenomena lead to production of high-energy particles and possibly for neutrino emission from high-energy particle interactions. The AGN source is at the location of the central supermassive black hole, while the starburst source is extended through the volume of the galaxy. The possibility of localizing the source and possibly resolving it would be helpful for distinguish different contributions to the neutrino flux in case of source detection.

Circinus is the one of the closest Type~2 Seyfert galaxies that has its central black hole of the mass comparable to that of the Milky Way black hole, $(1.7\pm 0.3)\times 10^6 M_\odot$ \citep{2003ApJ...590..162G}. Bolometric luminosity of the source reaches a sizable fraction of the Eddington luminosity for this black hole mass \citep{2026NatCo..17...42L}. The emission from the black hole is heavily obscured, making the source ``Compton thick'' AGN \citep{Matt:1998fr,2026NatCo..17...42L}, with the observed hard X-ray flux much below the Eddington limit. The Compton thick nature of the source explains large uncertainty in the estimate of the neutrino flux from the hard X-ray luminosity of the source that has to be reconstructed from modelling, rather than directly measured.  Signatures of particle acceleration in the source are revealed by Fermi/LAT detection \citep{2013ApJ...779..131H}. It is however not clear if the $\gamma$-ray signal originates from the AGN or starburst activity \citep{2019ApJ...885..117G}.  The source features a kiloparsec scale outflow that may be a jet originating from the AGN source \citep{10.1046/j.1365-8711.1998.01592.x}. An excess of neutrino events from the direction of the source in the  ``starting track'' event  has been reported by IceCube~\citep{2602.10208}. The source was found to contribute to an excess of neutrino counts from the direction of a set of X-ray bright AGN sample in the IceCube analysis. The significance of this excess for the individual source at the location of Circinus galaxy is below $3\sigma$, so that no evidence for the signal from this source has been claimed. Nevertheless, the IceCube paper provides an estimate of the source flux that is shown in Fig. \ref{fig:south_sources}. One can see that this flux estimate is consistent with the assumed scaling of the neutrino flux with the hard X-ray luminosity, being at the upper end of the range of source flux predictions shown in Fig. \ref{fig:south_sources}. If the neutrino flux from the source is indeed at the level estimated from the IceCube data, the source is expected to be readily detectable by KM3NeT, on the time scale of just several years at $3\sigma$ significance and in less than 10 years at $5\sigma$ significance, see Table \ref{tab:results}. 

NGC~4945 at the distance of just 3.8 Mpc is the brightest Seyfert galaxy in the sky in the 10-100 keV energy range \citep{1996ApJ...463L..63D}. It also hosts a black hole of the mass comparable to that of the Milky Way, $M_{BH}=3.4^{+3.2}_{-1.6}\times 10^6 M_\odot$ \citep{2015ApJ...802...98M}. It is yet another nearby AGN with bolometric luminosity that can be close to the Eddington limit \citep{2012MNRAS.423.3360Y}. Similar to Circinus and NGC 1068, it is also heavily obscured in the X-ray band.  An estimate of neutrino flux from this source suffers from the same difficulty as Circinus: the hard X-ray luminosity has to be deduced from spectral modeling, it is not directly measured. Moreover, an additional uncertainty is introduced by the variability of the source flux in hard X-ray range, by at least a factor of two between different NuSTAR observations \citep{2014ApJ...793...26P}. It is also a Fermi/LAT source \citep{2010A&A...524A..72L} showing signs of particle acceleration either through its AGN or starburst activity. This source has been included in the analysis of Southern hemisphere X-ray bright AGN by \citet{2602.10208}, but it is not associated to an excess in the data. This suggests that its neutrino flux is lower than that of Circinus galaxy and it possibly does not reach the highest level listed in Table \ref{tab:galaxies}. However, no definitive conclusions can be driven from the low statistics IceCube signal: the absence of excess can also be a statistical fluctuation. 

Our estimates of the neutrino luminosities of the two additional sources are based on the hypothesis of scaling of the source luminosity with the hard X-ray luminosity, presumably generated in a compact hot corona wrapping the central black hole and the innermost part of the accretion flow. This hypothesis needs to be tested and, in the case of its validity, the scaling between the neutrino and hard X-ray luminosities needs to be calibrated based on a large number of detected sources. This will be difficult even with KM3NeT or with the combined IceCube and KM3NeT source sample. Such calibration should be ideally done using a sample of Type~1 Seyfert galaxies where the central AGN source is not obscured in hard X-ray band. Only one out of four sources detectable with KM3NeT is Type~1 Seyfert galaxy, NGC~4151. Moreover, there is only one other Type~1 Seyfert galaxy, IC~4329A that may possibly be detectable if its spectrum is somewhat harder than the templates used in our study.  The obscuration of the central hard X-ray source in other sources introduces a large uncertainty in the estimates of the ``intrinsic'' hard X-ray luminosity of the source. 

Overall, we find that KM3NeT has the potential for independent verification of the IceCube claim of existence of neutrino signal from Seyfert galaxies via detection of at least two additional sources in the Southern hemisphere. It should even be able to localize the neutrino emission site within the source host galaxies. Extension of the neutrino-emitting Seyfert galaxy source sample will only provide a limited possibility for testing the hard-X-ray -- neutrino flux scaling hypothesis, because of heavy obscuaration of the X-ray emission in the two additional sources. 

\section{Acknowledgement}
The work of Mykhailo Poleshchuk was supported by the National Research Foundation of Ukraine under project 2023.03/0149.

\bibliographystyle{elsarticle-harv} 
\bibliography{refs}

\end{document}